\documentclass[11pt,dvips]{article}

\usepackage{epsfig,times}
\usepackage{picinpar}
\usepackage{wrapfig}
\usepackage{floatflt}
 \makeatletter
\renewcommand{\section}{\@startsection{section}{1}{0in}
    {0.4\baselineskip}{0.1\baselineskip}{\Large\bf}}
\renewcommand{\subsection}{\@startsection{subsection}{2}{0in}
    {0.25\baselineskip}{-\baselineskip}{\large\bf}}
\renewcommand{\subsubsection}{\@startsection{subsubsection}{3}{0in}
    {0.1\baselineskip}{-\baselineskip}{\normalsize\bf}}
\makeatother

\begin{document}
\makeatletter\newcommand{\ps@icrc}{
\renewcommand{\@oddhead}{\slshape{OG.3.2.30}\hfil}}
\makeatother\thispagestyle{icrc}
\begin{center}

{\LARGE \bf Ultra high energy neutrinos scattering off relic light
neutrinos to explain UHECR above GZK cut off and thin blazars
jets}
\end{center}

\begin{center}
{\bf D. Fargion $^1$, B. Mele $^1$}\\ {\it $^{1}$ Physics
Department, Rome University 1, and INFN, Rome1, P.za Aldo Moro 2
Rome, ITALY}
\end{center}

\begin{center}
{\large \bf Abstract\\}
\end{center}
\vspace{-0.5ex} UHE neutrinos may transfer highest cosmic-rays
energies overcoming $2.75K^\circ$ BBR and radio-waves opacities
(the GZK cut off) from most distant AGN sources at the age of the
Universe. These UHE $\nu$ might scatter onto those (light and
cosmological) relic neutrinos clustered around our galactic halo
or nearby neutrino hot dark halo clustered around the AGN blazar
and its jets. The branched chain reactions from a primordial
nucleon (via photo production of pions and decay to UHE neutrinos)
toward the consequent beam dump scattering on galactic relic
neutrinos is at least three order of magnitude more efficient than
any known neutrino interactions with Earth atmosphere or direct
nucleon propagation. Therefore the rarest cosmic rays (as the 320
EeV event) might be originated at far $(\tilde{>} 100 Mpc)$
distances (as Seyfert galaxy MCG 8-11-11); its corresponding UHE
radiation power is in agreement with the observed one in MeV gamma
energies. The final chain products observed on Earth by the Fly's
Eye and AGASA detectors might be mainly neutron and anti-neutrons
and delayed, protons and anti-protons at symmetric off-axis
angles. These hadronic products are most probably secondaries of
$W^+ W^-$ or $ZZ$ pair productions and might be consistent with
the last AGASA discoveries of  doublets and one triplet event.

\vspace{1ex}

\section{Introduction:}
\bigskip
 Most energetic cosmic rays UHE $(E_{CR} > 10^{19} eV)$ are
bounded to short $(\tilde{<} 10 Mpc)$ distances by the 2.73 K° BBR
opacity (Greisen K.1966; Zat'sepin G.T., Kuz'min V.A., 1966) (the
GZK cut off) and by diffused radio noise (Clark T.A. et
al.1970,Protheroe R.J.,Biermann P.L. 1996. The main
electromagnetic ``viscosities'' stopping the UHE cosmic ray
(nuclei, nucleons, photons, electrons) propagation above $\sim 10
Mpc$ are:

\noindent  (1) The Inverse Compton scattering of any charged
lepton (mainly electrons) on the BBR $(e^\pm_{CR} \gamma_{BBR}
\rightarrow e^\pm \gamma)$ (Longair 1994),(Fargion et all
1997),(Fargion D.,Salis A. 1998),

\noindent (2) the nucleons photopair production at higher energies
$(p_{CR} + \gamma_{BBR} \rightarrow p e^+ e^-)$,

\noindent (3) the UHE photon BBR or radio photon electron-pair
productions $(\gamma_{CR} \gamma_{BBR} \rightarrow e^+ e^-)$,

\noindent (4) nuclei fragmentation by photo-pion interactions.

\noindent (5) the dominant nucleon photo-production of pions
$(p_{CR} + \gamma_{BBR} ~\rightarrow ~ p + N \pi ; n +
\gamma_{BBR} \rightarrow n + N \pi)$.

The above GZK constrains apply to all known particles (protons,
neutrons, photons, nuclei) excluding neutrinos. Nevertheless, the
UHE cosmic rays, either charged or neutral, flight straight
keeping memory of the primordial source direction, because of the
extreme magnetic rigidity (Bird D.J. 1994). However all last well
localized and most energetic cosmic rays (as the Fly's Eye 320 EeV
event on October 1991) do not exhibit any cosmic nearby $(< 60
Mpc)$ source candidate in the same arrival direction error box.
Indeed even most recent data from AGASA (Takeda et all 1999)show
the clustering of doublets and one triplet UHE cosmic rays. These
events might be associated with Markarian sources (as Mrk 359,Mrk
40,Mrk 171) at distances (>50Mpc) well above GZK cut-off. From
here the UHE cosmic ray paradox arises: the few known solutions
are difficult to be accepted.
\section{Possible solution to GZK puzzle.}
\bigskip
\noindent (a) An {\it exceptional} $(B \tilde{>} 10^{-7} Gauss)$
{\it coherent} magnetic field on huge extra-galactic distance able
to bend (by a large angle) the UHECR trajectory coming not from
distant but from nearest off-axis sources like M 82 or Virgo A
(Elbert J.W,Somers P.,1995).  This solution was found not
plausible ( Medina-Tanco, G.A., 1997).

\noindent (b) Exotic topological defect annihilations (Elbert
J.W,Somers P.,1995) in diffused galactic halo is an ad hoc, and a
posteriori solution. Moreover it is in contradiction with recent
evidences by AGASA detector of data inhomogeneities, i.e. of
doublets or triplet UHECR events arriving from the same
directions. (Takeda M.  et al., 1998)

\noindent (c) A galactic halo population of UHECR sources (as the
fast running pulsars associated with SGRs(Fargion D., Salis A.,
1995 )). These small size (neutron star - black-hole) jets source
must be extremely efficient in cosmic ray acceleration at energies
well above the expected common maximum energy $E < BR \tilde{<}
10^{17} eV ({B\over 3.10^{-6} G}) ({R\over 50pc})$ required by a
supernova accelerating blast wave. Moreover their extended halo
distribution must exhibit a dipole and/or  a quadrupole UHECR
anisotropy, signature not yet identified. Therefore it seems at
least premature to call for a solution by local galactic or local
group sources as microquasar jets.

\noindent (d) A direct nucleus or a nucleon at high energies will
be severely suppressed by GZK $(10^{-5})$ cut off and, more
dramatically, they will induce a strong signal (hundreds) of
secondaries  cosmic rays at energies just below the GZK bound;
indeed such a copious signal is absent.
\section{UHE neutrino-antineutrino interaction in galactic halo solving
the GZK puzzle}
\bigskip
Our present (Fargion D., Salis A., 1997; (Fargion
D., Mele B., Salis A., 1997 ) solution is based on the key role of
light $(m_\nu \tilde{>} eV)$ cosmic neutrinos clustered in
extended galactic halo. These relic neutrinos act as a target
calorimeter able to absorb the UHE$\nu$'s from cosmic distances
and to produce hadronic showers in our galaxy. The primary UHECRs
are the usual AGNs or Blazars able to produce huge powers and
energies. Their electronic production and decays, near the source,
into muonic and electronic neutrinos, generate the main $\nu$s
messenger toward cosmic distance up to our galactic halo. Their
final interactions with clustered relic $\nu_r$ (and
$\bar{\nu}_r$) of all flavours (but preferentially with the
heaviest and best clustered one $(\nu_\tau , \bar{\nu}_\tau)$) may
offer different channel reactions:

\noindent (A) $\nu \bar{\nu}_r$ scattering via a Z exchanged in
the  s channel leading to nucleons and photons.

\noindent (B) $\nu \bar{\nu}_r$ scattering via t-channel of
virtual W exchange among different flavour. This is able to
produce copious UHE photons (mainly by $\nu_\mu \bar{\nu}_\tau
\rightarrow \mu^- \tau^+$ and $\tau$ pion decay),

\noindent (C) $\nu \bar{\nu}_r$ production of $W^- W^+$ or ZZ
pairs. The latter channels are in our opinion the best ones  to
produce final nucleons $(p, \bar{p} , n ,\bar{n})$ which fit
observational data.

\vskip 12pt \noindent {\bf The UHE $\nu$ cross section interacting
with relic $\nu_\tau, \bar{\nu}_\tau$.} \vskip 6pt

The general framework to solve the GZK puzzle we proposed is a
tale story beginning from a far AGN source whose UHE protons $(E_p
\tilde{>} 10^{23} eV)$ are themselves a source of pions,
secondaries muons and UHE neutrinos. The latter may actually
escape the GZK cut off, traveling unbounded all the needed cosmic
distances $(\sim 100 Mpc)$. Once near our galactic halo, the
denser gravitationally-clustered light relic neutrinos, forming a
hot dark halo, might be able to convert the UHE $\nu$'s energies
by scattering and subsequent decays into  observable nucleons (or
anti-nucleons), the final observed UHECR remnants. The $\nu - \nu$
interaction cross-sections are the key filter which makes possible
and efficient the whole process.  In fig. 1 from (Fargion D., Mele
B., Salis A., 1997) we show the three main processes cross-section
as a function of the center of mass energy s. The s - channel
$\nu_\mu \bar{\nu}_{\mu R} \rightarrow Z$ which exhibits a
resonance at $E_\nu = 10^{21} eV ({m\nu\over 4eV})^{-1}$ and the t
- channel $\nu_\mu \nu_{\tau_R} \rightarrow \mu^- \tau^+$ via
virtual W. These reactions are the most probable ones but UHE
photons seem to be excluded by geomagnetic high altitude cut off.
The $\nu \bar{\nu}_\tau \rightarrow W^+ W^-$ cross section is also
shown. There is an additional Z pair production channel $\nu
\bar{\nu}_\tau \rightarrow ZZ$  almost coincident in its general
behaviour with the  $W^+ W^-$ production. It is not shown on the
figure, but its global contribution (to be discussed elsewhere) is
to almost double the $ \nu \bar{\nu}_\tau \rightarrow W^- W^+$
chain products making easier their detection. The reaction chain
from the primordial proton, via electronic production and
neutrino-neutrino interactions  down to the final cosmic ray event
considered, the corresponding probability the consequent
multiplicity are discussed in  (Fargion D., Mele B., Salis A.,
1997).

Assuming that the clustered neutrino density contrast is
comparable to the barionic one ${n\nu_\tau\over n \nu_{BBR}} \sim
{\rho_G\over \rho_W} \sim 10^{5\div 7}$, one finds the total
probability of the processes and the corresponding needed
primordial proton energy $E_p^{WW}$. The probability (taking into
account also the global multiplicity ) to occur is at least
$P^{WW} \tilde{>} 10^{-3} ({m\nu\over 10eV})^{-1}$ corresponding,
for the candidate source MCG 8-11-11, to a needed average power
$E^{WW} \sim 1.2 \cdot 10^{48} ergs^{-1}$. This value  is
comparable with the MCG8-11-11 observed low-gamma MeV luminosity
$(L_\gamma \sim 7 \cdot 10^{46} erg s^{-1})$. We predicted
parasite signals photons at $10^{16} eV$ energies  (Fargion D.,
Mele B., Salis A., 1997) as well as a peculiar imprint on larger
sample data, due to the central overlapping of neutron-antineutron
prompt arrival toward the source line of sight. Additional twin
(deviated) signals due to proton and antiprotons random walk will
arrive late nearly at opposite (few degrees off-axis) sides. These
characteristic signatures might be already recorded by AGASA in
the last few doublet and triplet UHECR and associated events. The
UHE neutrinos above the GZK cut-off are observable from almost all
the Universe while the corresponding UHE nucleons (or gamma) above
the GZK energies are born in a smaller constrained ``GZK'' volume
of a ten of Mpc. Therefore the expected flux ratio for UHE $\nu$'s
over nucleons or photons at GZK energies is roughly (in Euclidean
approximation) independently on the source spectra:

\[{\phi_\nu \over \phi_{GZK}} \sim [{z_\nu \over z_{GZK}}]^{3/2}
\simeq 3 \cdot 10^4 \eta ({z_\nu\over 2})^{3/2} ({z_{GzK}\over 2
\cdot 10^{-3}})^{-3/2}\]

\noindent
 where $z_\nu$ is a
characteristic UHE $\nu'$ source redshift $\simeq 2$ and $z_{GZK}
\sim 2\cdot 10^{-3}$. Assuming an efficiency  ratio $\eta$ for the
conversion from UHE proton to UHE $\nu$'s of a few percent, the
ratio $\phi_\nu /\phi_{GZK} \tilde{>} 10^3$ is naturally
consistent with the inverse probability $(P^{WW} \tilde{>}
10^{-3})^{-1}$  found above.

Therefore a $10^{+3}$ fold larger flux of UHE $\nu$'s (than the
corresponding nucleon cosmic ray flux ) above GZK (mainly of tau
nature (Fargion D., B.Mele, Salis A., 1997) should be observed
easily in a $Km^3$ neutrino detector in a very near future. This
prediction is somehow in disagreement with recent controversial
upper bounds on neutrino fluxes (Waxmann \& Bachall 1999).

Finally the same presence of relic clustered neutrinos around the
AGN source should reveal the thin neutrino jet at wide distances
($\sim$Mpc) from central engine of the jet. Indeed the UHE $\nu$
jet may dissipate (with low but non negligible efficiency) its
energy with the relic $\nu$s at far distances ($\geq 10 Kpc$)
leading to UHE electron pairs which may be the source (by
synchrotron radiation) of X and gamma radiation all along the thin
jet. These UHE electrons, if they had directly arisen from the
source, cannot escape far away (few pc) from the AGN because of
the Inverse Compton Scattering ``viscosity''. Their presence in
the thin AGN jet far away (tens Kpc) from the source might be
associated only with UHE $\nu$-$\nu_r$ scattering as proposed in
the  present paper.
\begin{center}
{\Large\bf References}
\end{center}
Greisen K., 1966, Phys. Rev. Lett., 16, 748\\ Zat'sepin G.T.,
Kuz'min V.A., 1966, JETP Lett, 4, 78\\ Clark T.A., Brown L.W.,
Alexander J.K., 1970, Nature, 228, 847\\ Protheroe R.S., Biermann
P.L., 1996, Astropart. Phys. 6, 45\\ Longair M.S., 1994, Hygh
Energy Astrophysics, Vol, 2, Cambridge University Press. and
references therein.\\ Fargion D., Konoplich R.V., Salis A., 1997,
Z.Phys. C, 74,571\\ Fargion D., Salis A., 1998, Phys. Uspeki,
41(8), 823\\ Bird D.J. et al., 1994, ApJ, 424,491\\ Elbert J.W.,
Somers P., 1995, ApJ, 441, 151\\ Medina-Tanco, G.A., 1997,
astro-ph/9610171\\ Fargion D., Salis A., 1995, Nuclear Physics B,
43, 269-273\\ Takeda M. et al., 1998, Phys. Rev. Lett. 81,
1163-1166 and Astro-ph/9807193\\ Fargion D., Salis A., 1997, Proc.
25th ICRC, Patchetstroomse, South Africa, 7, HE4-6, p157\\ Fargion
D., B.Mele, Salis A., 1997, astro-ph/9710029; ApJ, 1999, 517,
725\\ Zel'dovich, Ya B., 1980, Sov. J.Nucl. Phys., 31, 664\\
Fargion D., 1997, astro-ph/9704205;1999 submitted ApJ.\\Takeda M.
et al., 1999,astro-ph/9902239.
\vspace{1ex}
\begin{figwindow}[1,r,%
{\mbox{\epsfig{file=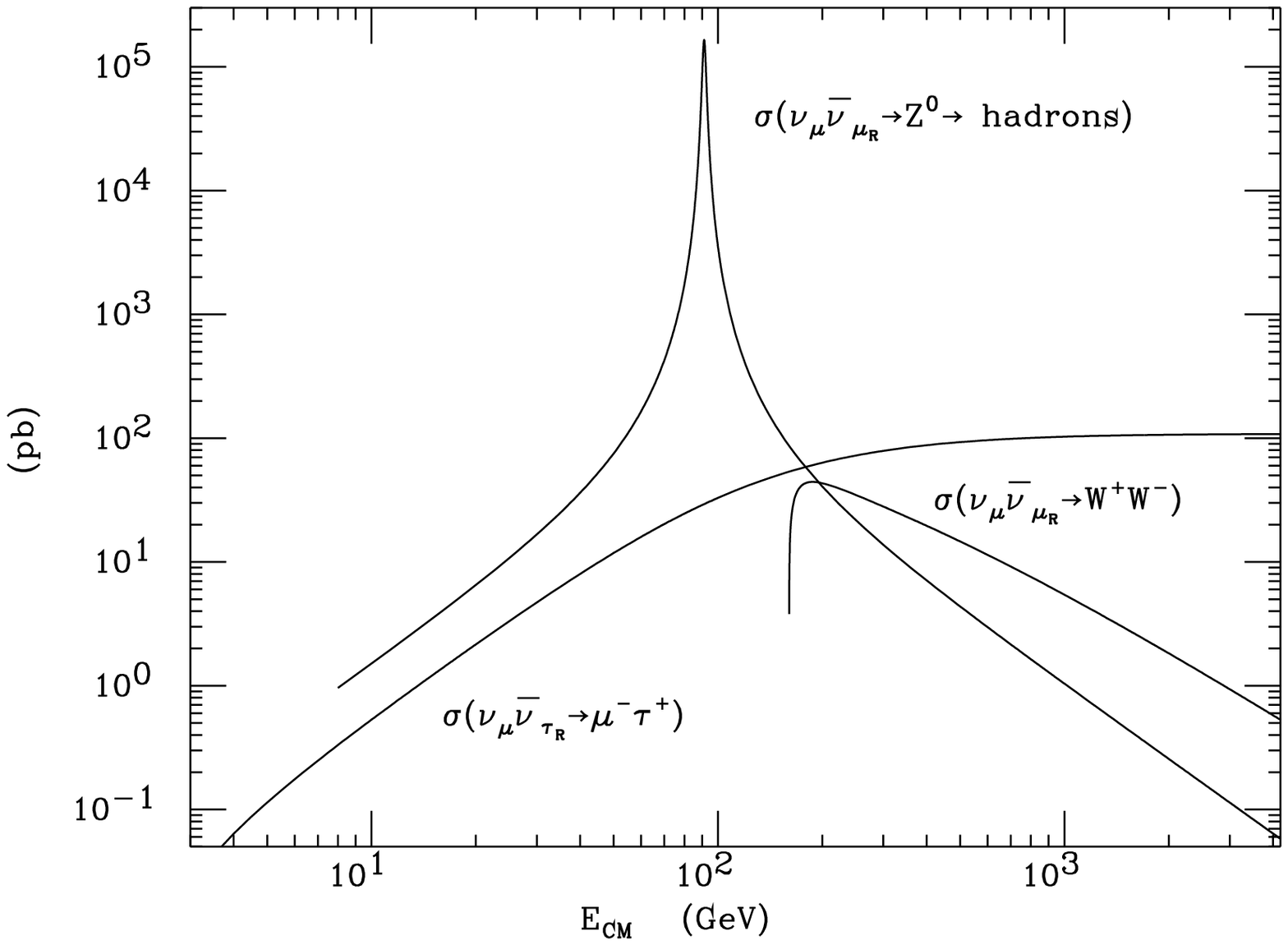,}}},%
{}]
\end{figwindow}

\end{document}